\documentclass[12pt,a4paper,final]{iopart}
\usepackage{iopams}  
\usepackage{graphicx}
\usepackage[breaklinks=true,colorlinks=true,linkcolor=blue,urlcolor=blue,citecolor=blue]{hyperref}

\newcommand{\vc}{\mathbf}
\usepackage{graphicx}
\usepackage{color}

\begin{document}


\title[Instability-Enhanced Friction in Two-Ion Species Plasmas]{Instability-Enhanced Friction in the Presheath of Two-Ion-Species Plasmas}


\author{Scott D.\ Baalrud}
\address{Department of Physics and Astronomy, University of Iowa, Iowa City, Iowa 52242, USA}

\author{Trevor Lafleur$^{1,2}$}
\address{$^1$Laboratoire de Physique des Plasmas, CNRS, Sorbonne Universit\'{e}s, UPMC Univ Paris 06, Univ Paris-Sud, Ecole Polytechnique, 91128 Palaiseau, France}
\address{$^2$ONERA-The French Aerospace Lab, 91120 Palaiseau, France}

\author{William Fox}
\address{Princeton Plasma Physics Laboratory, Princeton, New Jersey 08543, USA}

\author{Kai Germaschewski}
\address{Space Science Center, University of New Hampshire, Durham, New Hampshire 03824, USA}



\begin{abstract}

The speed at which ions enter a sheath is a fundamental property of a plasma that also provides a useful boundary condition in modeling. A recent theory proposed that this can be significantly influenced by an instability-enhanced friction force arising from two-stream instabilities in the presheath when multiple ion species are present. Although experiments appeared to confirm this theory, recent particle simulations have brought it into question. We reconcile this controversy using direct numerical solutions of the dispersion relation, which show that there is a dependence on the electron-ion temperature ratio that was not considered previously. In addition, particle-in-cell simulations are used to show that ion-ion two-stream instabilities can arise near the sheath edge and generate an enhanced ion-ion friction force. Only by accounting for the instability-enhanced friction force can theory predict the simulated ion speeds at the sheath edge.

\end{abstract}



\vspace{2pc}
\submitto{\PSST}


\maketitle


\section{Introduction}

Knowing the speed at which ions leave a plasma is important for accurate modeling of both plasma-material interactions and global plasma characteristics. This affects laboratory plasmas, particularly in the areas of diagnostics and materials processing~\cite{lieb:05}, as well as the interaction of space plasmas with spacecraft~\cite{robe:13,garr:81}. Langmuir~\cite{tonk:29} and Bohm~\cite{bohm:49} first predicted that ions leave a plasma supersonically when a single ion species is present: $V_i \geq c_s \equiv \sqrt{T_e/M_i}$, where $T_e$ is the electron temperature and $M_i$ is the ion mass. This constraint is called the Bohm criterion and it has long been agreed upon theoretically~\cite{riem:91}, and confirmed experimentally \cite{goec:92,oksu:02}, that the equality form is satisfied in quiescent weakly collisional plasmas. However, there is currently a lack of consensus between theory, experiments and simulations on how to determine the speed of each ion species if more than one is present~\cite{fran:00,seve:03,hers:05,hers:05b,wang:06,lee:07,oksu:08,baal:09,baal:11,yip:10,hers:11,gudm:11,hers:12,seve:13}. This is a critical issue because plasmas often contain a mixture of ion species and the speed of each is desired as a boundary condition. 

The Bohm criterion has been extended to include multiple ion species \cite{cook:80,riem:95}. For two ion species the result is
\begin{equation}
\frac{n_1}{n_e} \frac{c_{s1}^2}{V_1^2} + \frac{n_2}{n_e} \frac{c_{s2}^2}{V_2^2} \leq 1 \label{eq:bohm2}
\end{equation}
where the indices 1 and 2 label the ion species and $c_{si} = \sqrt{T_e/M_i}$ is a sound speed associated with species $i$. We assume $T_i \ll T_e$ in Eq.~(\ref{eq:bohm2}). Like the single species Bohm criterion, equality is expected to hold in Eq.~(\ref{eq:bohm2}) (this has also been confirmed experimentally~\cite{hers:11}). However, an additional constraint is required to determine both $V_1$ and $V_2$ at the sheath edge. If the two ion species are decoupled, the presheath electric field may be expected to impart the same energy to each. Applying $M_1 V_1^2/2 = M_2 V_2^2/2$ to Eq.~(\ref{eq:bohm2}) leads to the prediction that each species obtains its individual sound speed at the sheath edge ($V_i = c_{si}$). Ion-neutral drag causes only slight deviations from this prediction in weakly collisional plasmas~\cite{fran:00}. In the other limit, complete coupling between the two ion species, such that $V_1=V_2$ in Eq.~(\ref{eq:bohm2}), predicts flow at a common system sound speed $c_{s} = \sqrt{(n_1/n_e) c_{s1}^2 + (n_2/n_e) c_{s2}^2}$. Although there was apparent theoretical consensus on the individual sound speed solution~\cite{lieb:05,riem:95}, experiments in Ar$^+$-Xe$^+$ plasmas measured values closer to the system sound speed~\cite{seve:03,hers:05,hers:05b,wang:06,lee:07,oksu:08}. This merging of ion speeds suggests that ion-ion friction may be playing a role in the experiments, but standard Coulomb collisions are far too weak to explain the measurements~\cite{baal:09,baal:11}.

A recent theory proposed that ion-ion two-stream instabilities in the presheath can cause an instability-enhanced friction (IEF) force through wave-particle scattering~\cite{baal:09,baal:11}. This theory is consistent with the experimentally-observed flow coupling, and experimental evidence for ion-ion two-stream instabilities using probes has also been presented~\cite{hers:05,yip:10,hers:11}. Although experiments appeared to provide verification~\cite{yip:10,hers:11}, recent particle-in-cell/Monte Carlo collision (PIC-MCC) simulations by Gudmundsson and Lieberman (GL) were shown to disagree with it~\cite{gudm:11}. These simulations found no evidence of the instabilities, or enhanced friction force, which challenged the theory and raised the question of why it agrees with the experiments but not the simulations. PIC-MCC is a widely used simulation technique providing one of the most detailed descriptions of plasma kinetics, so resolving this discrepancy is a critical issue. 

Here, we reconcile this apparent discrepancy. First, using exact numerical solutions of the linear dispersion relation, we show that there is a critical temperature ratio required for instability $T_e/T_i \geq (T_e/T_i)_c$ that was not considered previously. We show that while the temperature ratio from the GL simulation \cite{gudm:11} falls below this threshold, the exact theory still predicts instability for the parameters of the LHS experiment~\cite{lee:07}. Second, and more importantly, we conduct new PIC-MCC simulations that confirm the presence of ion-ion two-stream instabilities near the sheath edge. We find that the ion-ion friction force is enhanced in this same region. Furthermore, the simulated ion speeds agree well with the values predicted by the IEF theory. 

Two-stream instabilities arise when the differential ion flow speed exceeds a threshold condition: $V_1 - V_2 > \Delta V_c$.\footnote{We adopt the convention that species 1 is less massive than species 2, so $V_1 > V_2$.} Differential ion flow arises as the presheath electric field accelerates the lighter species to a faster speed (e.g., within the ballistic presheath model $V_i =c_{si}$ at the sheath edge). The IEF force is predicted to prevent the differential flow from significantly exceeding the threshold condition ($V_1 - V_2 \simeq \Delta V_c$). However, if $\Delta V_c > |c_{s1} - c_{s2}|$ ions reach the sheath edge without exciting instabilities, so the traditional individual sound speed solution is expected to hold. Thus, the condition 
\begin{equation}
\Delta V \equiv V_1 - V_2 = \min \lbrace \Delta V_c, |c_{s1} - c_{s2}| \rbrace  \label{eq:v12crit}
\end{equation}
has been proposed as a second constraint that along with Eq.~(\ref{eq:bohm2}) provides both $V_1$ and $V_2$. Two analytic approximations have been provided for $\Delta V_c$ based on the value of the ion mass ratio~\cite{baal:09,baal:11}. These predict that there is a concentration dependence on the ion speeds, which has been tested experimentally~\cite{yip:10,hers:11}.  Here we go beyond these approximations to show that there is also a temperature ratio threshold for instability. 

This paper is organized as follows. Section~\ref{sec:linear} describes numerical solutions of the full kinetic linear dispersion relation for ion-ion two-stream instabilities. Limitations of the previous analytic approximations are discussed. Particular emphasis is placed on the instability thresholds in differential flow speed ($\Delta V_c$) and temperature ratio $(T_e/T_i)_c$. The latter condition was not considered in the previous approximate solutions. Section~\ref{sec:pic} describes the results of PIC-MCC simulations of He-Xe plasmas at different concentrations. This provides a number of original results: (1) The first identification of ion-ion two-stream instabilities in the presheath in a particle simulation, (2) the first identification of an enhancement of the ion-ion friction force directly correlated with these instabilities, and (3) the first simulation test of the ion flow speeds predicted by the IEF theory in a situation where these instabilities are present. 

\section{Linear Instability Thresholds~\label{sec:linear}} 

\subsection{Linear Kinetic Theory} 

This section advances the previous analytic approximations for the instability threshold conditions~\cite{baal:11} by treating numerical solutions of the full kinetic linear dispersion relation. To do so, we consider the standard linear dielectric response function for ion-frequency fluctuations in an unmagnetized plasma~\cite{swan:03}
\begin{equation}
\hat{\varepsilon} = 1 + \frac{1}{k^2 \lambda_{De}^2} \biggl[ 1 - \frac{1}{2} \frac{T_e}{T_1} \frac{n_1}{n_e} Z^\prime (\xi_1) - \frac{1}{2} \frac{T_e}{T_2} \frac{n_2}{n_e} Z^\prime (\xi_2) \biggr]  \label{eq:ephat}
\end{equation}
where $\xi_1 = \hat{k} \cdot \Delta \vc{V} (\Omega -1/2)/v_{T1}$, $\xi_2 = \hat{k} \cdot \Delta \vc{V} (\Omega + 1/2)/v_{T2}$, $Z^\prime (\xi) = dZ/d\xi$ and $Z$ is the plasma dispersion function. The parameter $\Omega$ has been defined by the substitution 
\begin{equation}
\omega = \frac{1}{2} \vc{k} \cdot (\vc{V}_1 + \vc{V}_2) + \vc{k} \cdot \Delta \vc{V} \Omega , \label{eq:osub}
\end{equation}
where $\omega$ is the complex angular wave frequency. This substitution highlights the fact that the growth rate depends on $\Delta \vc{V}$ and that the most unstable wavenumber is parallel to $\Delta \vc{V}$. The assumptions leading to Eq.~(\ref{eq:ephat}) are that ions have Maxwellian distribution functions, unity charge, frequencies of interest satisfy $\omega/kv_{Te} \ll 1$, and that the gradient scale length of plasma parameters (densities, temperatures and flow speeds) are much longer than the wavelengths of interest. The latter holds because the Debye-scale waves of interest are much shorter than the presheath. Ion-neutral collisions are neglected.  We also assume that the two ion species have the same temperature $T_1 = T_2 =T_i$.\footnote{In our simulations we find that the He$^+$ and Xe$^+$ temperatures are typically within 10\% of each other. }

\begin{figure}
\begin{center}
\includegraphics[width=8cm]{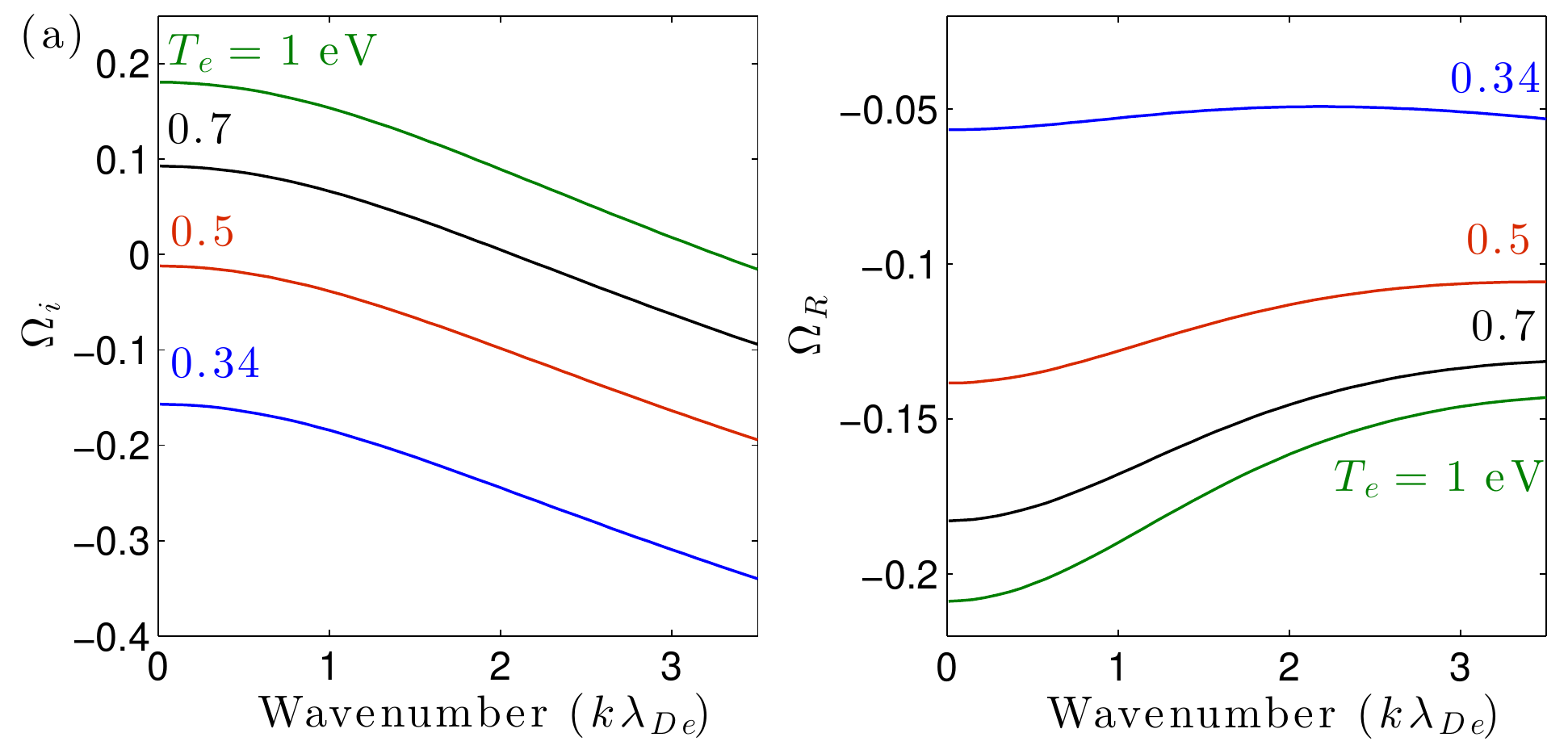}\\
\includegraphics[width=8cm]{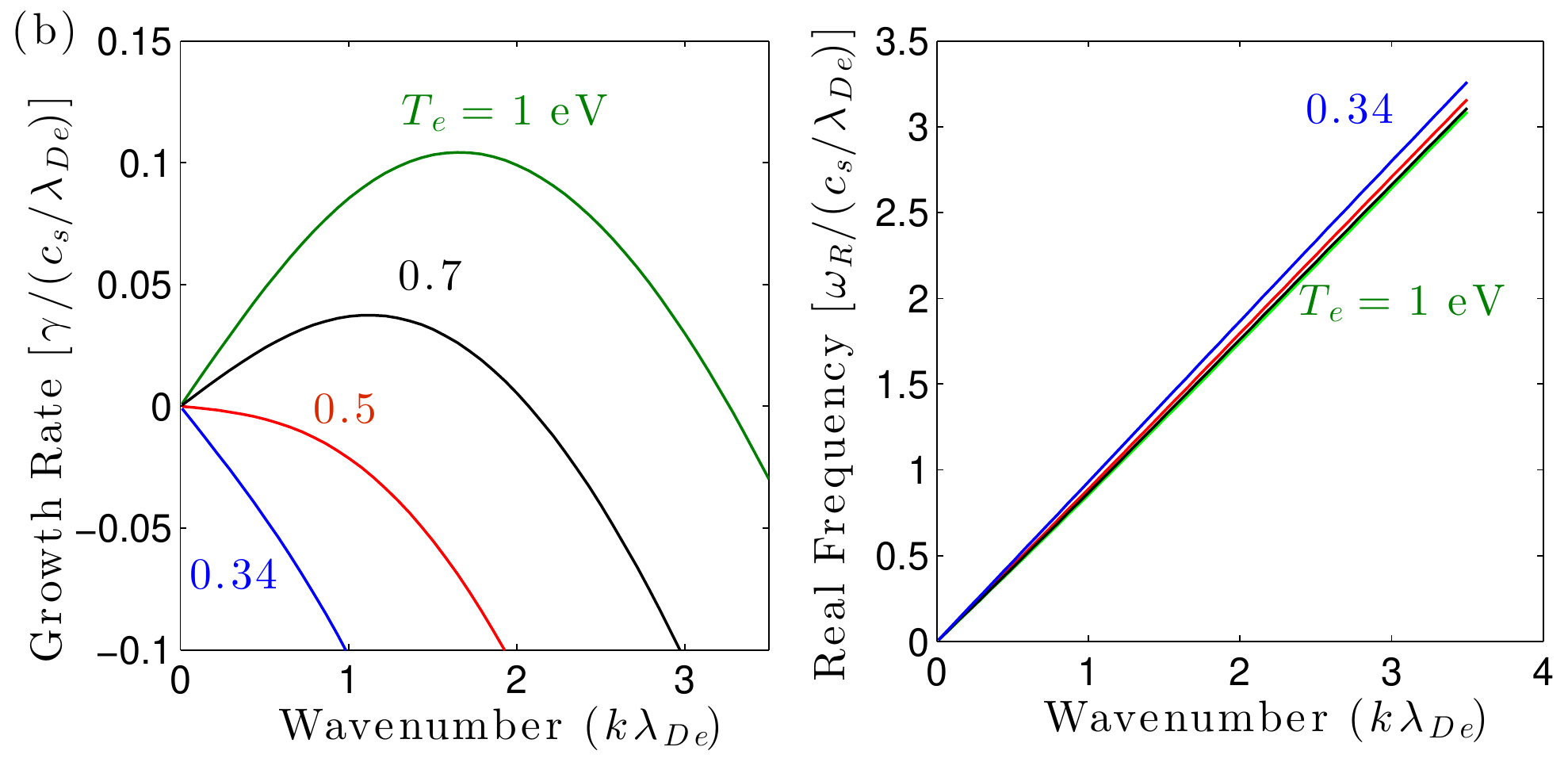}\\
\includegraphics[width=8cm]{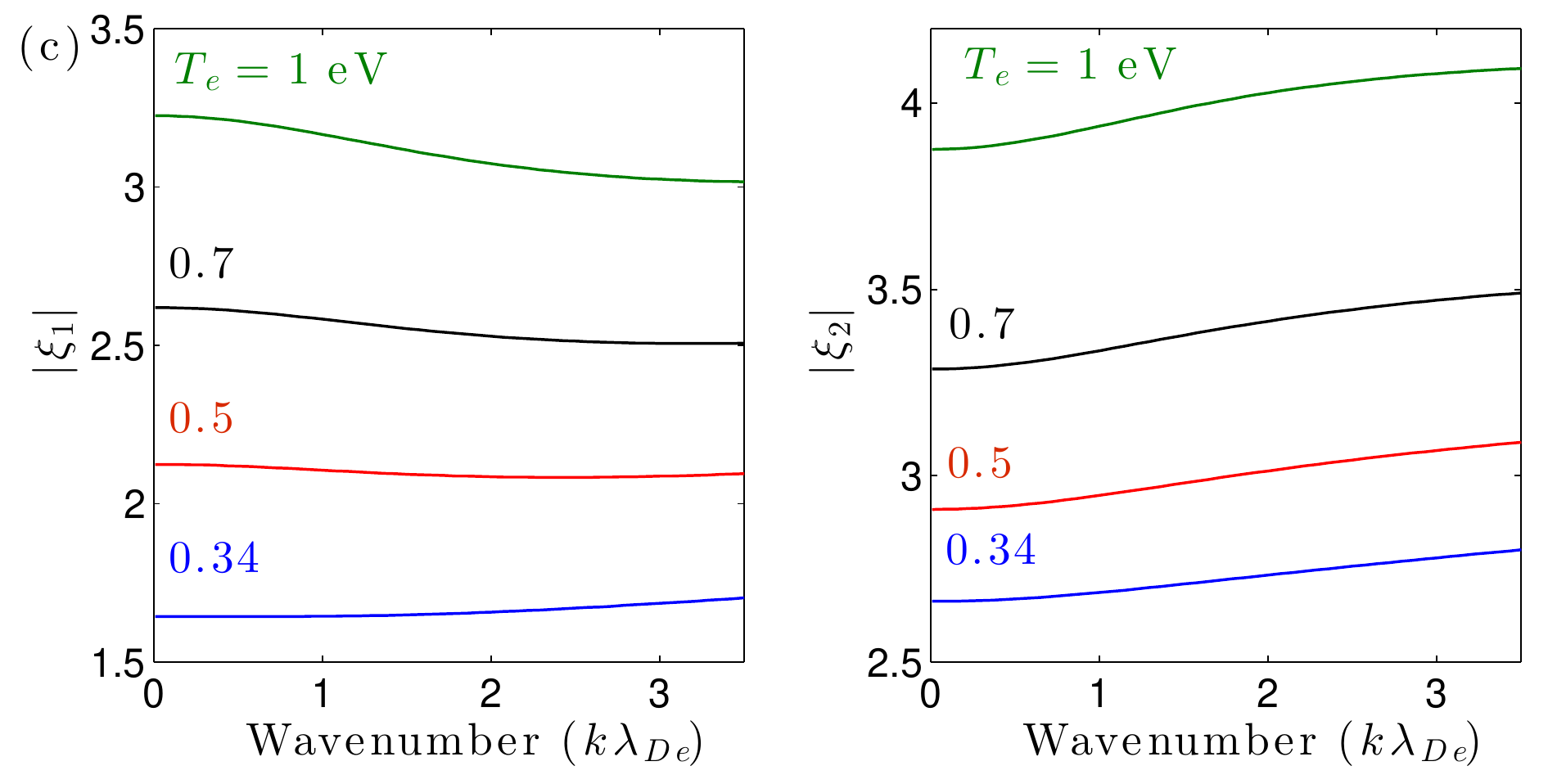}
\caption{(a) The parameters $\Omega_I$ and $\Omega_R$ in a Ar-Xe plasma using four values of the electron temperature $T_e = 1, 0.7, 0.5$ and $0.34$ eV. The differential flow was taken to be $\Delta V = c_{s1} - c_{s2} = 692, 579, 490$ and $404$ m/s at each electron temperature. The other parameters used were $T_i = 0.023$ eV and $n_1/n_e = 0.5$.  (b) Growth rates and real frequency corresponding to the solutions from (a). (c) Magnitude of the arguments of the $Z$-functions in Eq.~(\ref{eq:ephat}). }
\label{fg:growth_ar}
\end{center}
\end{figure}

\begin{figure}
\begin{center}
\includegraphics[width=8cm]{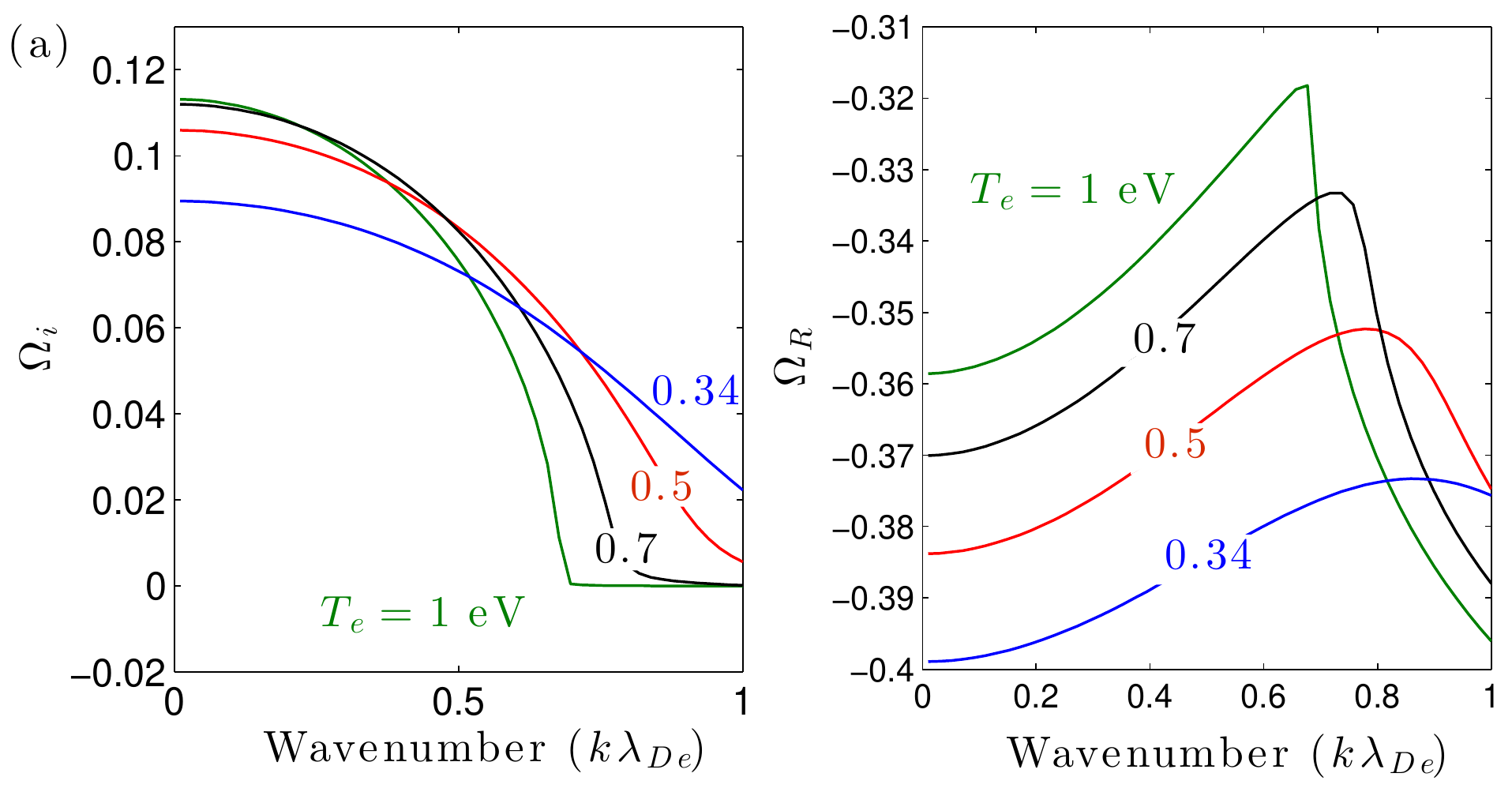}\\
\includegraphics[width=8cm]{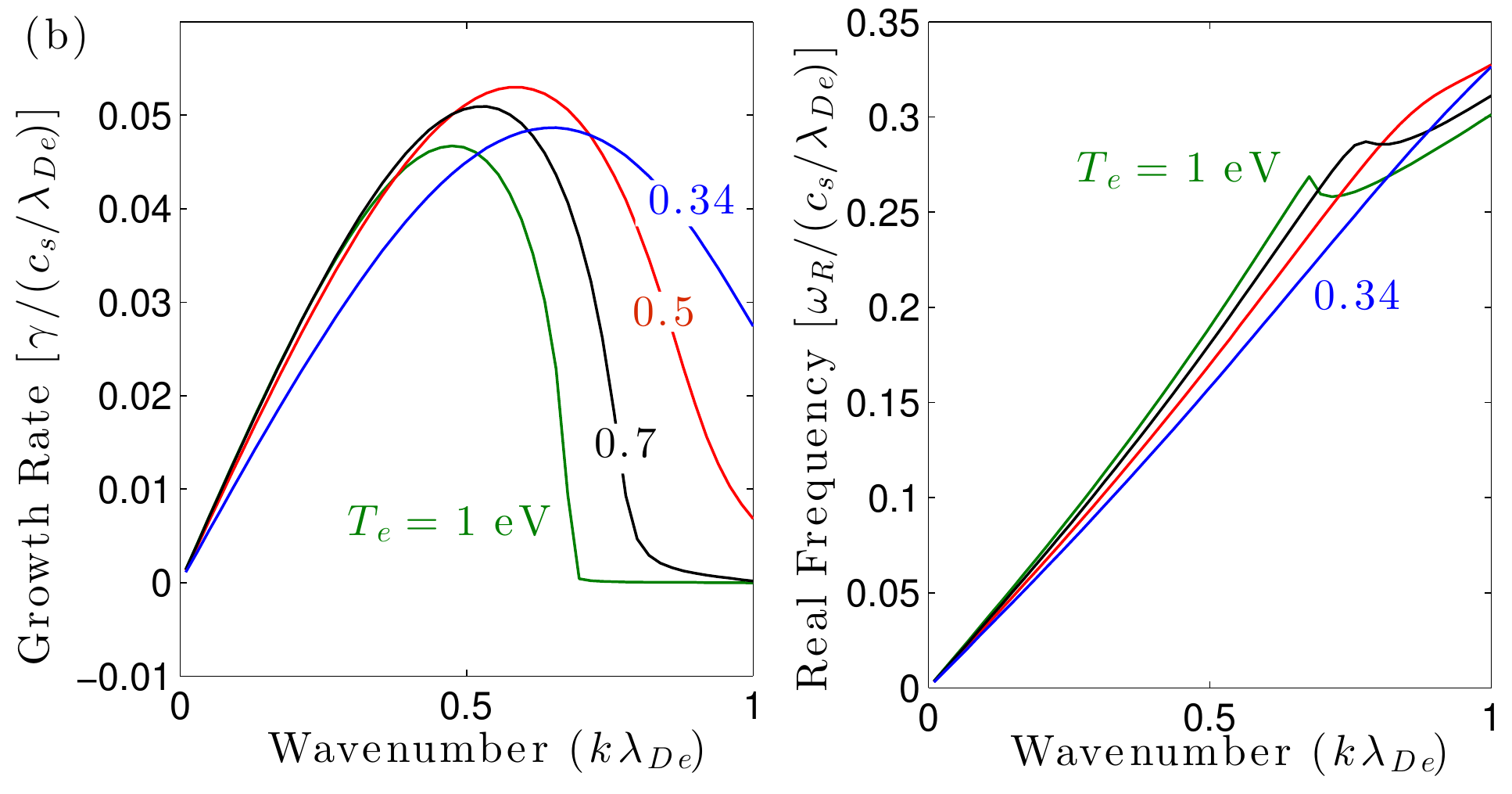}\\
\includegraphics[width=8cm]{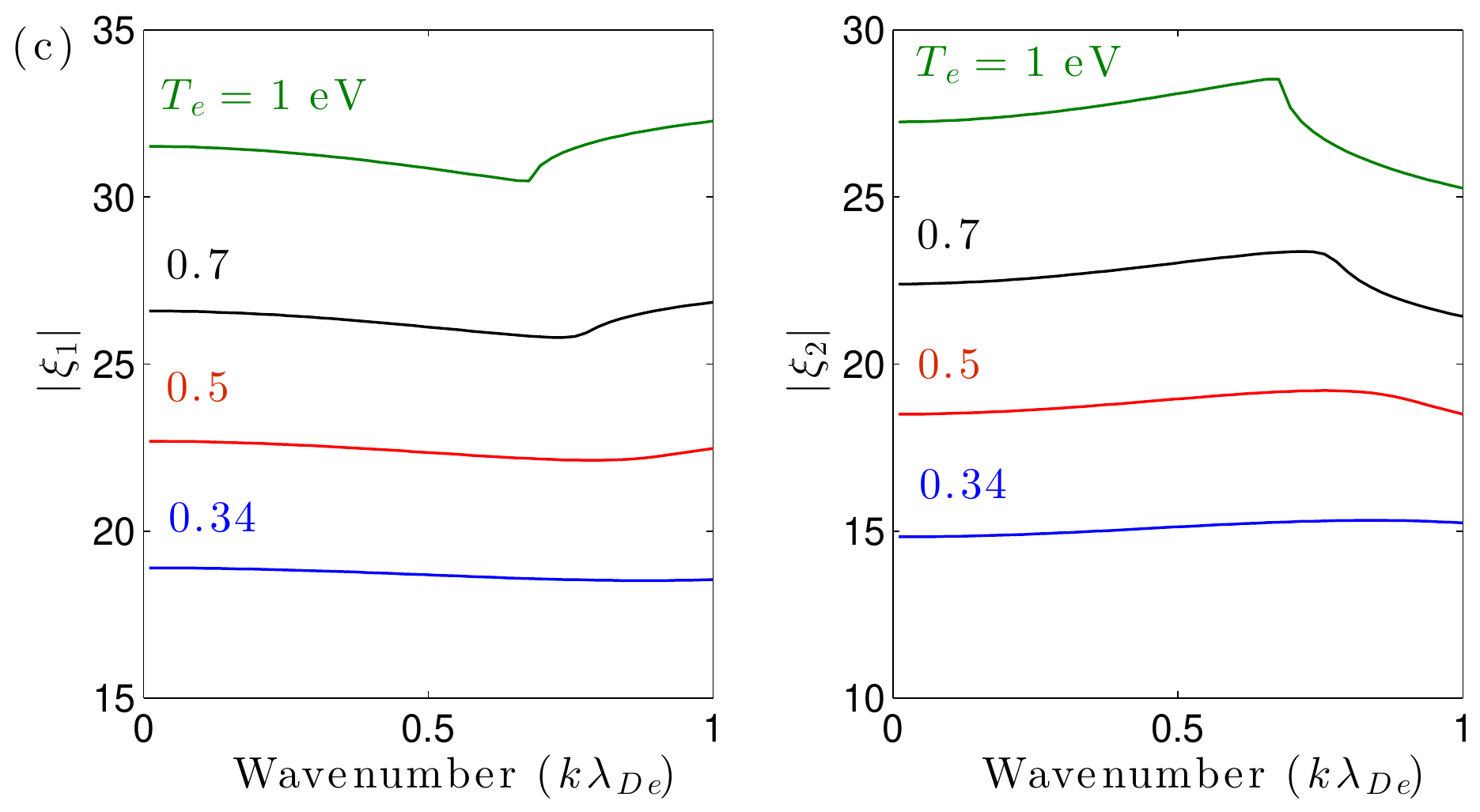}
\caption{(a) The parameters $\Omega_I$ and $\Omega_R$ in a He-Xe plasma using four values of the electron temperature $T_e = 1, 0.7, 0.5$ and $0.34$ eV. The differential flow was taken to be $\Delta V = c_{s1} - c_{s2} = 6.1, 5.1, 4.3$ and $3.5$ km/s at each electron temperature. The other parameters used were $T_i = 0.023$ eV and $n_1/n_e = 0.5$.  (b) Growth rates and real frequency corresponding to the solutions from (a). (c) Magnitude of the arguments of the $Z$-functions in Eq.~(\ref{eq:ephat}). }
\label{fg:growth_he}
\end{center}
\end{figure}

Solutions of the dispersion relation, $\hat{\varepsilon}(\omega,\vc{k})=0$, computed from Eq.~(\ref{eq:ephat}) are shown in Fig.~\ref{fg:growth_ar} for Ar-Xe plasma and in Fig.~\ref{fg:growth_he} for He-Xe plasma, each at typical parameters for low temperature plasma experiments (parameters given in the figure captions). One salient feature of the solutions is that the Ar-Xe growth rate becomes a purely damped solution below a modest electron temperature of approximately 0.5 eV. This indicates that there is a temperature ratio threshold that was not considered in \cite{baal:11}. It appears to be a possible cause of the discrepancy between the GL simulations~\cite{gudm:11} and LHS experiments~\cite{lee:07} (this point analyzed in more detail in Sec.~\ref{sec:tc}). 

Another important aspect is shown in panel (c) in each figure. This gives the argument of each of the $Z$-functions in Eq.~(\ref{eq:ephat}). Fluid treatments of the two-stream instability arise from a large argument expansion. The figures show that the arguments are not large in the Ar-Xe case, implying that a kinetic treatment is required. The arguments are large in the He-Xe case, justifying a fluid expansion, except when the electron temperature drops too low. Reference \cite{baal:11} discusses the kinetic versus fluid aspect at length, where it is suggested that the magnitude of these arguments depends on the mass ratio of ion species. For disparate masses $\sqrt{M_2/M_1} \gtrsim 4$ (or $\sqrt{M_2/M_1} \lesssim 1/4$) it was suggested that the normal fluid expansion should be accurate, providing an analytic approximation for $\Delta V_c$. For similar masses $1/4 \lesssim \sqrt{M_2/M_1} \lesssim 4$ it was suggested that a different (order unity argument) expansion of the $Z$-functions was required, which led to a different analytic approximation for $\Delta V_c$. The numerical analysis shown in Figs.~\ref{fg:growth_ar} and \ref{fg:growth_he} largely affirms this expectation. However, the previous approximations also assumed $T_e /T_i$ to be asymptotically large. The numerical analysis shows that kinetic effects onset when the temperature ratio drops below approximately 10. This implies that there is a temperature ratio threshold for instability, and that calculating this requires a kinetic treatment. This is discussed in Sec.~\ref{sec:tc}.

\begin{figure}
\begin{center}
\includegraphics[width=7cm]{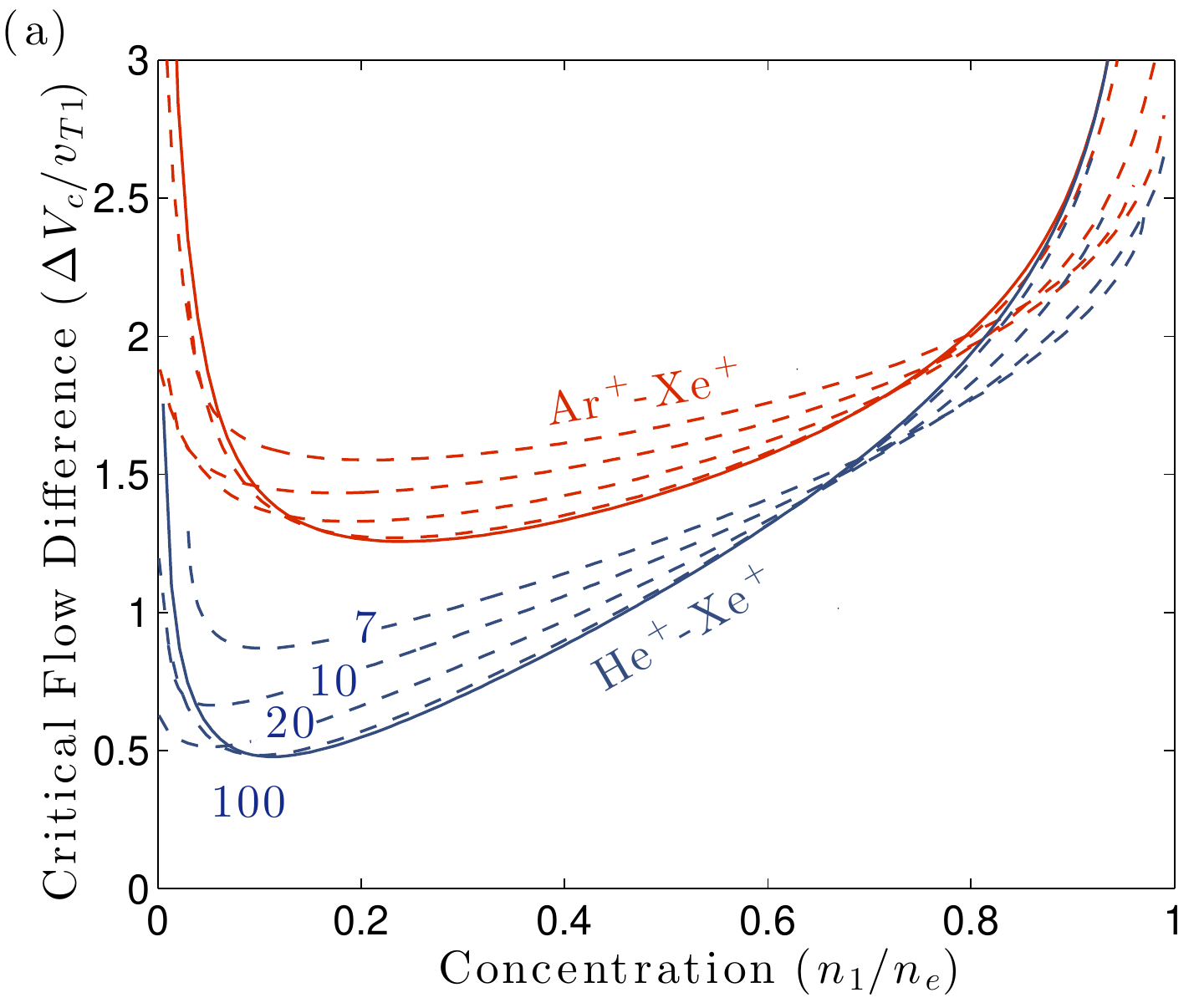}\\
\includegraphics[width=7cm]{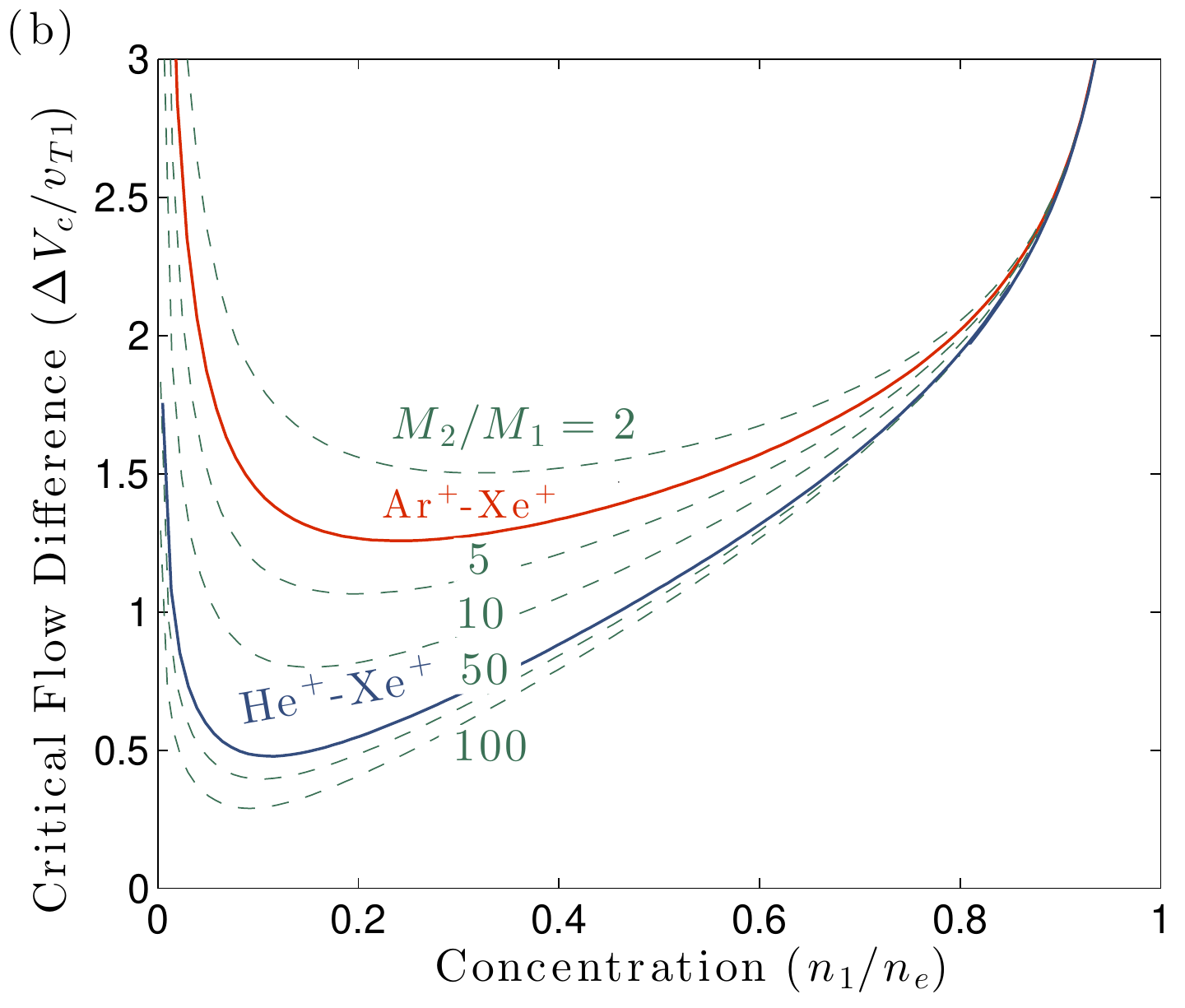}
\caption{(a) Critical flow difference for Ar$^+$-Xe$^+$ (red) and He$^+$-Xe$^+$ (blue) mixtures as a function of concentration. Dashed lines correspond to solving Eq.~(\ref{eq:f}) with $T_e/T_i = 7,10,20$ and 100, while the solid curves drop the third term in Eq.~(\ref{eq:f}).  (b) Critical flow difference for different ion mass ratios computed for large $T_e/T_i$, obtained by dropping the third term in Eq.~(\ref{eq:f}). }
\label{fg:dvc}
\end{center}
\end{figure}

\subsection{Instability Threshold: $\Delta V_c$}

The instability threshold in differential flow $\Delta V_c$ provides the input to the IEF theory consisting of Eqs.~(\ref{eq:bohm2}) and (\ref{eq:v12crit}). This can be calculated numerically from Eq.~(\ref{eq:ephat}) by varying the flow difference $\Delta V$ then finding the threshold value where the largest imaginary component of the dispersion relation first crosses the instability threshold [$\max \lbrace \omega_I (k, \Delta V)\rbrace = 0$]. For this calculation, Eq.~(\ref{eq:osub}) highlights a useful simplification; that $\max \lbrace \Omega_I (k) \rbrace$ occurs when $k = 0$. This is also shown in panel (a) of Figs.~\ref{fg:growth_ar} and \ref{fg:growth_he}. We emphasize that the maximum growth rate ($\gamma = \vc{k} \cdot \Delta \vc{V} \Omega_I$) is not at $k=0$ in general, but it approaches this value at the instability threshold. Thus, the threshold can be calculated by taking $k=0$ in Eq.~(\ref{eq:ephat}) without introducing any approximation. The threshold condition is then determined from
\begin{equation}
f(\Delta V_c) = c Z^\prime (\xi_1) + (1-c) Z^\prime (\xi_2) - 2 (T_i/T_e) = 0 .  \label{eq:f}
\end{equation}
where $c=n_1/n_e$ is the concentration of species 1. 

\begin{figure}
\begin{center}
\includegraphics[width=7cm]{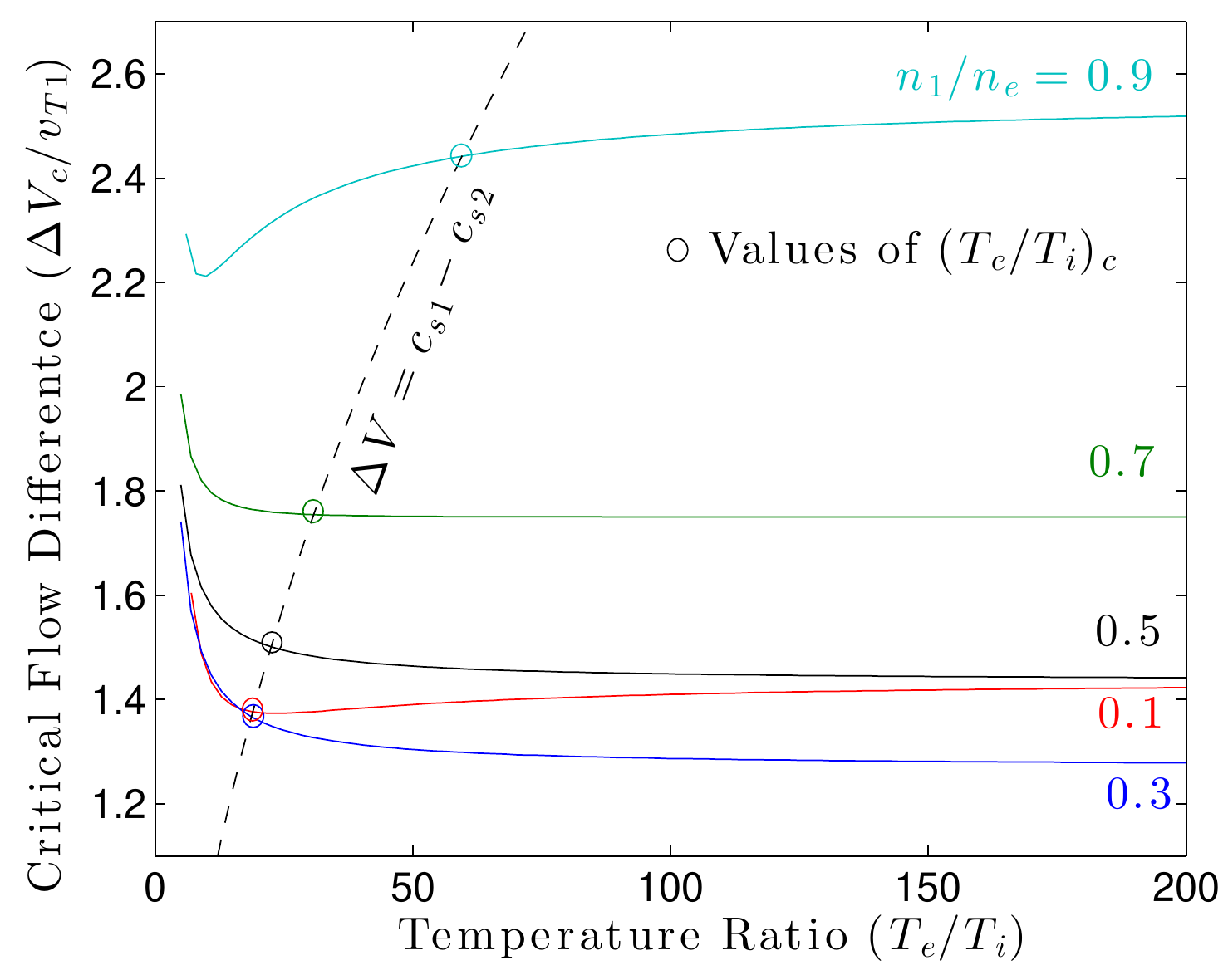}
\caption{Critical differential ion flow as a function of temperature ratio at several values of concentration for a Ar-Xe plasma. The curve for differential flow equal to the difference in individual sound speeds is also indicated. The intersection of this with the solid curves represents the critical temperature ratio for instability to onset in the presheath $(T_e/T_i)_c$.}
\label{fg:dvc_te}
\end{center}
\end{figure}

Solutions of $\Delta V_c$ computed from Eq.~(\ref{eq:f}) are shown in Fig.~\ref{fg:dvc} as a function of concentration. Panel (a) shows solutions for the Ar$^+$-Xe$^+$ and He$^+$-Xe$^+$ mass ratios at several values of the temperature ratio. These are the mixtures that have been studied experimentally \cite{seve:03,hers:05,hers:05b,wang:06,lee:07,yip:10,hers:11}. The previous approximate theories were based on the limit of asymptotically large $T_e/T_i$, removing the third term in Eq.~(\ref{eq:f})~\cite{baal:11}. Figure~\ref{fg:dvc} shows that this is accurate only when the temperature ratio is sufficiently large. Panel (b) shows solutions obtained in this large temperature ratio limit [by dropping the third term in Eq.~(\ref{eq:f})] at several different mass ratios. The normalized critical flow difference is lower at larger mass ratios. Figure~\ref{fg:dvc_te} shows solutions of the critical flow difference as a function of temperature ratio for Ar-Xe mass ratio at several concentrations. On one hand, this supports the notion that the temperature ratio dependence vanishes when it is sufficiently large. On the other hand, it shows that the temperature ratio has to be quite large in order for this to be valid at all concentrations.  

\begin{figure}
\begin{center}
\includegraphics[width=7cm]{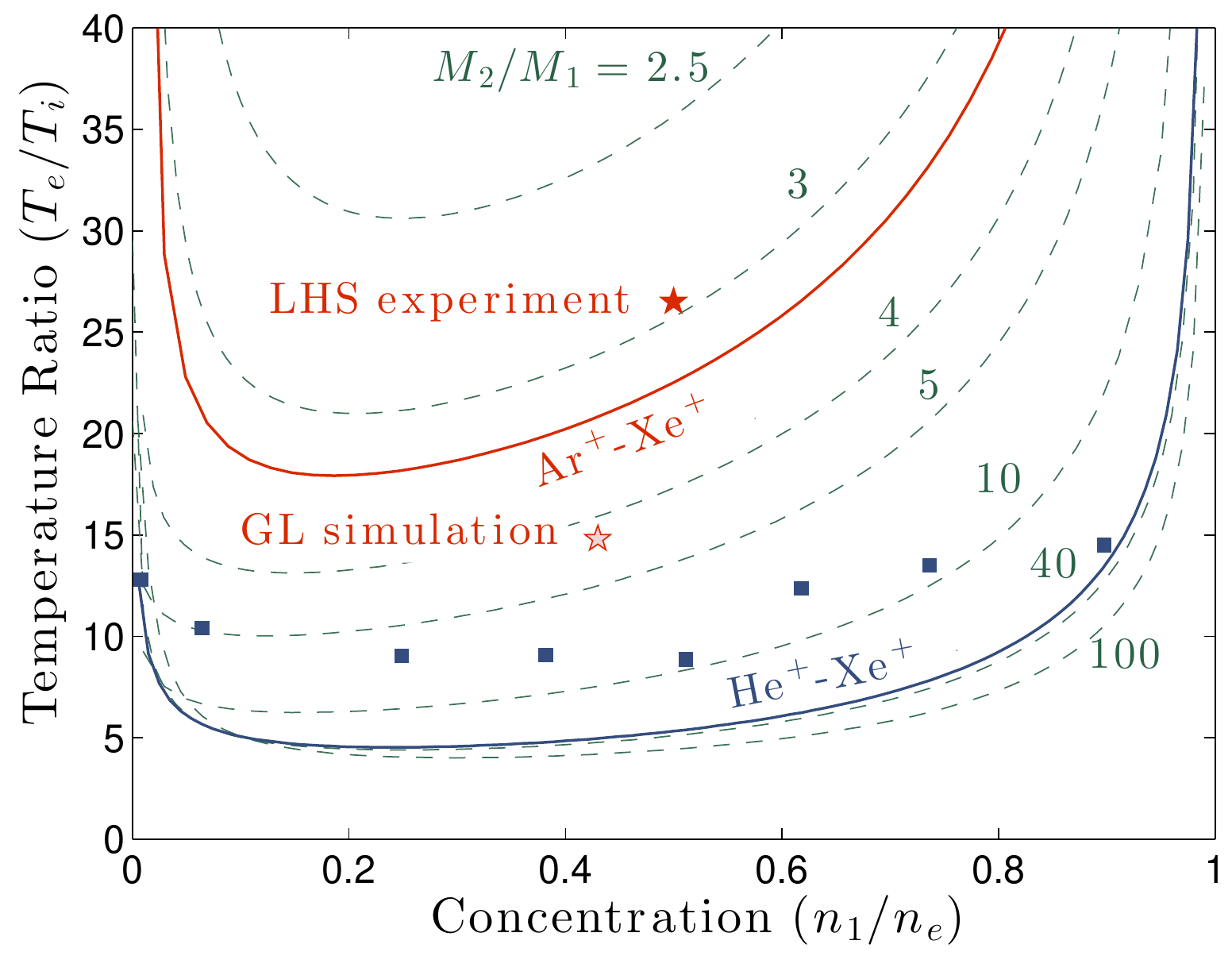}
\caption{Critical temperature ratio computed from Eq.~(\ref{eq:f}) as a function of concentration. Two-stream instability is predicted in the presheath for temperature ratios above the lines at each mass ratio. Also shown are the parameters for the LHS experiment~\cite{lee:07} and the GL simulation~\cite{gudm:11}. These were Ar$^+$-Xe$^+$ mixtures so the experiments fall in the unstable parameter space, but the simulations do not. Blue squares show the parameters of our new simulations of He$^+$-Xe$^+$ mixtures.}
\label{fg:tc}
\end{center}
\end{figure}

\subsection{Instability Threshold: $(T_e/T_i)_c$\label{sec:tc}}

The temperature ratio dependence can be further quantified by identifying a critical value required for instability in the presheath. Taking the differential flow speed to be the maximum possible value at the sheath edge, $\Delta V = |c_{s1} - c_{s2}|$, Eq.~(\ref{eq:f}) can be solved for the critical temperature ratio $(T_e/T_i)_c$ required for the IEF effect to arise in the presheath (if $\Delta V_c > |c_{s1} - c_{s2}|$ the entire presheath is stable). Making this substitution, the arguments of the $Z$-functions in Eq.~(\ref{eq:f}) depend only on $(T_e/T_i)_c$ and $M_1/M_2$. Figure~\ref{fg:dvc_te} overlays the $\Delta V = |c_{s1} - c_{s2}|$ curve on the critical flow difference plots for Ar-Xe. The intersection of these curves represent the critical temperature ratio points $(T_e/T_i)_c$. 

Figure \ref{fg:tc} shows solutions for this critical temperature ratio as a function of concentration at several ion mass ratios.  For temperature ratios below the lines, the presheath is predicted to be stable and each ion species is expected to reach its individual sound speed at the sheath edge. For temperature ratios above the lines, the presheath is predicted to be unstable and the speed of each ion species is expected to be determined from the solution of $V_1 - V_2 = \Delta V_c$ and Eq.~(\ref{eq:bohm2}). This figure shows that the LHS experiment~\cite{lee:07} falls in the unstable region of this phase-space, while the GL simulation~\cite{gudm:11} falls in the stable region. Thus, the reported discrepancy~\cite{gudm:11} can be explained by the assumption of an asymptotically large temperature ratio in the previous analytic approximations~\cite{baal:11}, which misses the temperature ratio threshold. The numerical solution of the instability bounds shows that the IEF theory is consistent with both the previous experiments and simulations.

\section{PIC-MCC Simulations~\label{sec:pic}}

\subsection{Code Description} 

Next, we present new PIC-MCC simulations to: (1) Search for the presence of two-stream instabilities in a regime where linear theory predicts that they should be present, (2) compare the ion-ion friction force in cases with and without instability, and (3) test the theoretical prediction that the differential flow is limited by the instability threshold. These were carried out using the {\sc phoenix1d} code previously described in~\cite{baal:11b}.  This electrostatic code simulated one spatial dimension, three velocity dimensions, and included a standard Monte Carlo routine to model particle-neutral collisions~\cite{vahe:95}. No explicit Coulomb collision routine was included. The simulation domain was 10 cm with grounded absorbing walls; secondary electron emission was not included. Electrons were heated by applying a sinusoidally varying electric field (with a frequency of 13.56 MHz) transverse to the simulation direction, and whose magnitude was determined self-consistently at each time step~\cite{meig:05}.  The control parameter in this scheme was the total transverse current, which is representative of that in an rf antenna~\cite{meig:05}. This heating scheme differs from that used in GL~\cite{gudm:11}, where particles were injected with an externally specified source term. The different schemes primarily affect the electron temperature.  Our PIC model is representative of a number of rf inductively coupled plasma (ICP) reactors~\cite{scim:98} and correctly models experimental pressure-domain length correlation~\cite{lieb:05}.

\begin{figure*}
\begin{center}
\includegraphics[width=5.1cm]{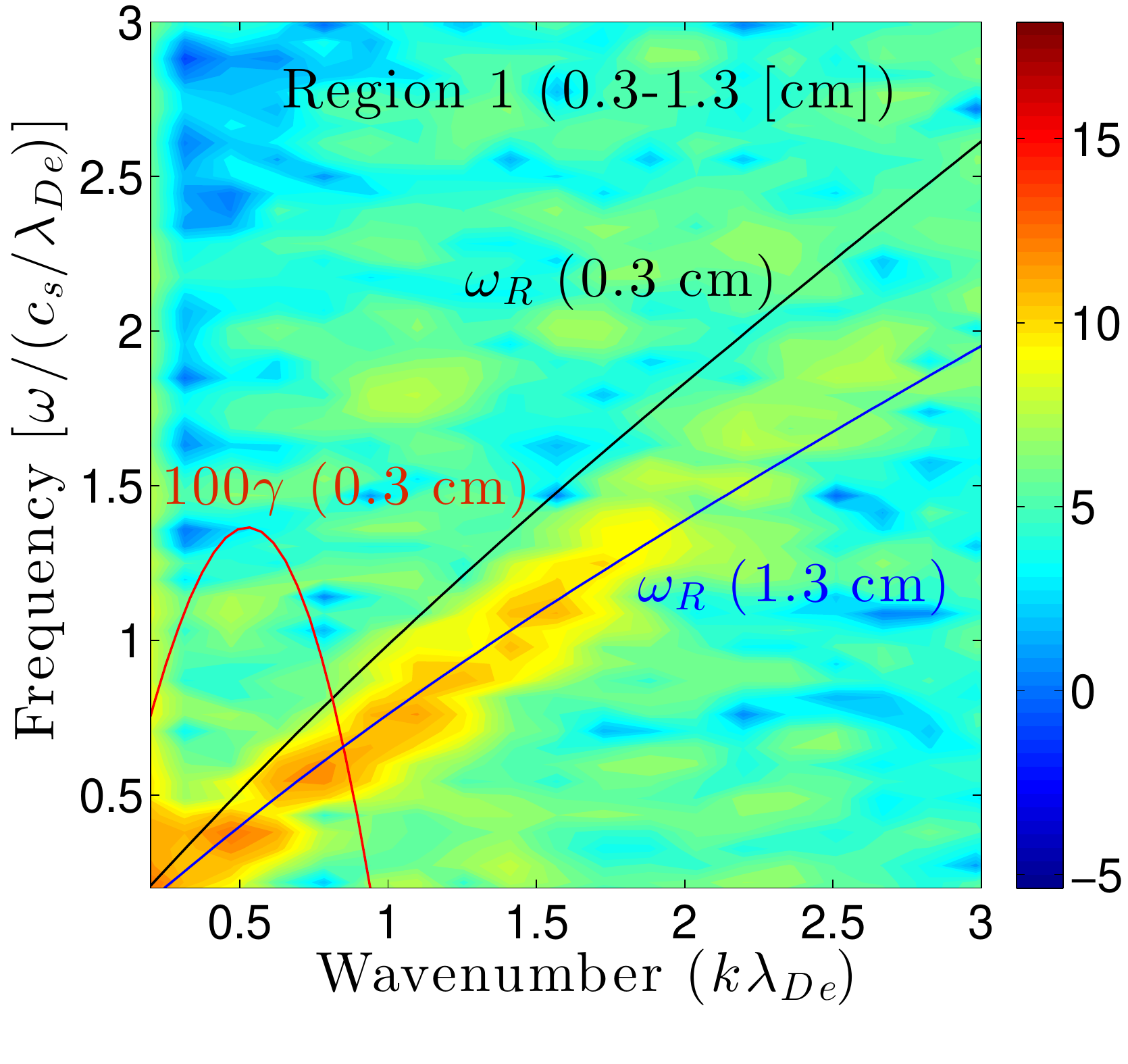}
\includegraphics[width=5.1cm]{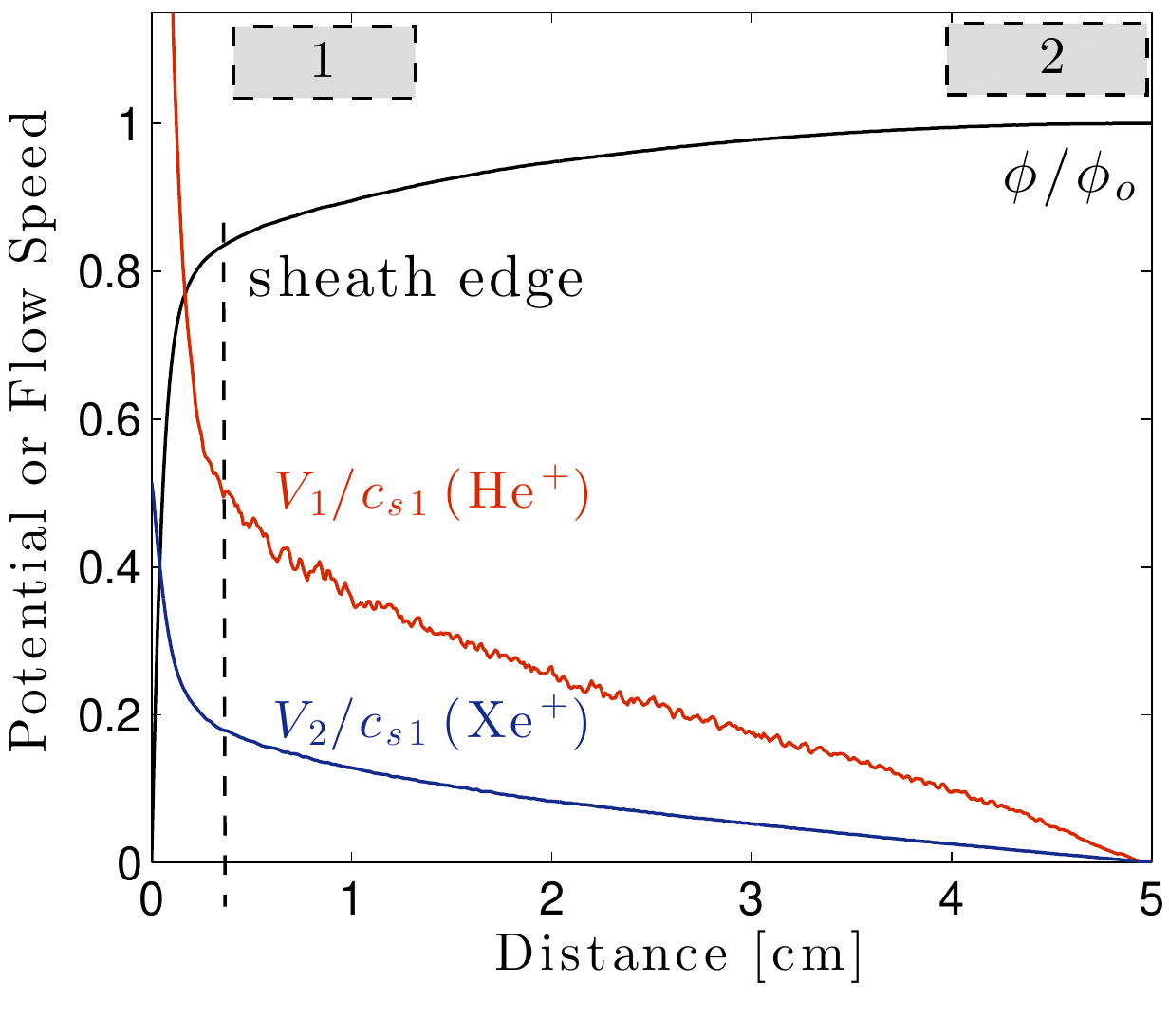}
\includegraphics[width=5.1cm]{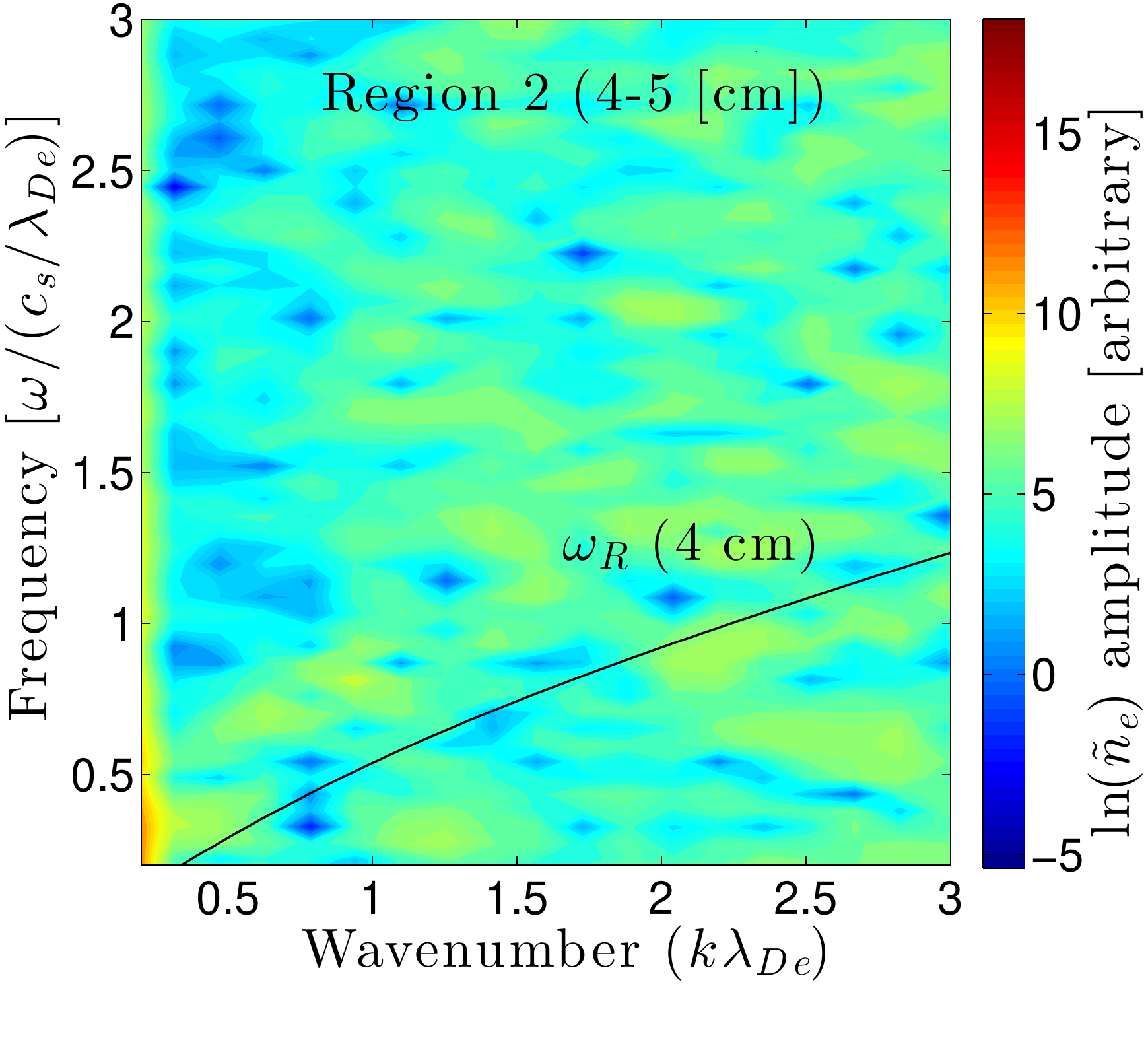}
\caption{ Fourier transform of electron density fluctuations near the sheath edge (region 1) and near the bulk (region 2) in a He$^+$-Xe$^+$ plasma. The middle panel shows the potential normalized by the central plasma potential (19.7 V) and the flow speed profiles. Regions 1 and 2 are indicated by the grey boxes. Normalizations use the indicated sound speeds based on the simulated electron temperature of $T_e = 4.3$ eV and central (5 cm) He$^+$ concentration of $n_1/n_e = 0.072$. }
\label{fg:fft}
\end{center}
\end{figure*}

We focus on He-Xe mixtures because the large disparity in ion masses leads to a lower temperature ratio threshold for instability; see Fig.~\ref{fg:tc}. In addition, all ion-neutral collisions were turned off to provide a cleaner test of the theory which also ignores these. Electron-neutral collisions for elastic scattering, excitation and ionization were included with cross-sections from Refs.~\cite{biag:04,phel:05,rapp:65,haya:83}. The He and Xe neutral gases were taken to be uniform with a constant temperature of 300~K at partial pressures ranging from 10-500 mTorr and 0.01-0.5 mTorr respectively. The large disparity in partial pressures was required to achieve the ion concentration range shown in the figure because the electron-impact ionization threshold is much lower for xenon than helium, and because the cross-sections are larger for xenon than helium. We used 1000 grid points, 2000 steps per rf cycle, and a few hundred thousand particles.  The ``antenna current'' control parameter was chosen to obtain maximum plasma densities of approximately $1 \times 10^{16}$~m$^{-3}$.  The simulations were run until steady-state, which typically took around 2000 rf periods. 

\subsection{Identification of Two-Stream Instabilities} 

Figure~\ref{fg:fft} shows an example of simulation results where ion-ion two-stream instabilities were observed in the presheath. The Fourier transform of electron density fluctuations in two domains is shown: the entrance to the presheath (4-5 cm), where no instabilities were observed, and near the sheath edge (0.3-1.3 cm), where a clear signature of ion-ion two-stream instabilities was observed. These enhanced fluctuations can be identified as ion-ion two-stream instabilities by comparing with the frequency predicted by the linear theory. Overlaid on this plot are the real frequency ($\omega_R$) and growth rate ($\gamma$) predicted by a numerical solution of the linear dispersion relation from Eq.~(\ref{eq:ephat}). These were calculated using characteristic temperatures from the center of the simulation domain $T_e = 4.3$ eV, $T_i=0.3$ eV, and the concentration and flow speeds from the simulation at 0.3, 1.3 and 4 cm. No two-stream mode exists at 5 cm because the ion flow is zero there.  

The observed fluctuation spectrum near the sheath edge agrees quite closely with the frequencies predicted by linear theory. Some frequency broadening is predicted from linear theory alone since the differential flow speed changes through the domain sampled by the FFT (this is the region between the two $\omega_R$ curves in the figure). However, some of the frequency broadening may also be an indicator of nonlinear effects. The range of unstable wavenumbers observed extends beyond that predicted by the linear theory (by approximately a factor of 2). One possible explanation is that this stems from inaccuracies of the Maxwellian approximation for the ion distribution functions. Figure~\ref{fg:growth_he} shows that the range of unstable wavenumbers depends on the electron-ion temperature ratio, which may effectively change due to deviations from Maxwellians. Here, the temperatures were obtained from moments of the distribution functions. It is worth noting that the dispersion relation is particularly influenced by the region of ion phase-space resonant with the mode.  Another possibility is that this too is an indicator of nonlinear features in the fluctuations (characteristic of a cascade to smaller scales). 

Depending on the He-Xe concentration levels, tests showed that the formation of instabilities were consistently observed, and tended to be insensitive to the number of grid points or number of particles used, so long as the instability wavelength was resolved.  Because ICP systems show electron distribution functions that vary with spatial position~\cite{meig:06} due to high-energy electron losses at the boundaries, tests were also performed using ``artificial'' electron-neutral cross-sections for helium, set numerically equal to those for xenon.  This confirmed that no false instabilities were observed due to differing spatial ionization rates of the two different ion species.


\subsection{Identification of Instability-Enhanced Friction} 

\begin{figure}
\begin{center}
\includegraphics[width=8cm]{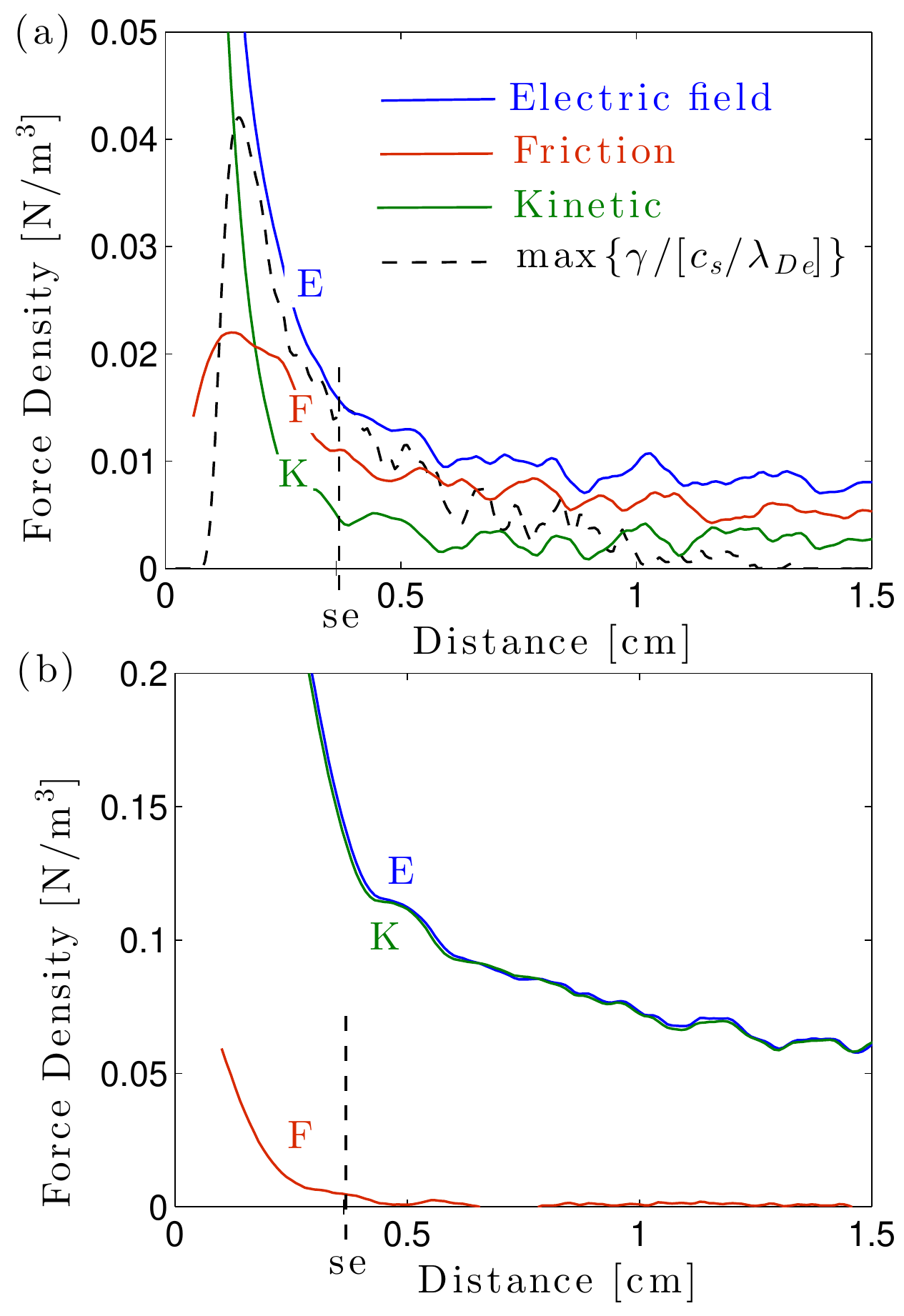}
\caption{Magnitude of each term in the He$^+$ momentum balance equation: Electric field force density ($-n_1 q_1 E$), friction force density ($R_{1}$) and kinetic force density  ($-\frac{d}{dx} \int d^3v m_1 v_x^2 f_1$). (a) At concentration $n_1/n_e = 0.072$, which was a concentration at which two-stream instabilities were observed (see Fig.~\ref{fg:fft}). The maximum linear growth rate calculated from Eq.~(\ref{eq:ephat}) is also shown for this concentration. (b) At concentration $n_1/n_e = 0.9$, which was a concentration at which two-stream instabilities were not observed. }
\label{fg:force}
\end{center}
\end{figure}

The ion flow speed profiles in Fig.~\ref{fg:fft} show that He$^+$ ions are approximately 50\% slower than their individual sound speed at the sheath edge, in accordance with the theoretical prediction at this concentration; see Fig.~\ref{fg:compare}. Figure~\ref{fg:force} shows the profile of the magnitude of each term in the time-averaged He$^+$ momentum balance equation (the momentum moment of the kinetic equation):
\begin{equation}
\frac{d}{dx} \biggl( \int d^3v m_1 v_x^2 f_1 \biggr) - n_1q_1 E = R_1 . \label{eq:hemom}
\end{equation}
Here the friction force is the drag felt by ions, which would otherwise respond ballistically to the electric field force. It results from collisions $R_1 = \int d^3v m_1 v_x C_1$, where $C_1$ is the Helium collision operator, and may include contributions from wave-particle collisions when instabilities are present~\cite{baal:10}. We do not calculate $R_1$ directly in the simulations, but rather infer this from the residual of the terms on the left side of Eq.~(\ref{eq:hemom}).  The 1D version is considered because the simulations have one spatial dimension. 

The terms on the left are the force densities associated with the kinetic energy of particles and electric field, respectively. These were calculated directly from the simulation data.  The data shown were averaged over 200 rf periods for a total of $4\times 10^6$ timesteps, and smoothed over 7 grid cells (0.7 mm).  Panel (a) shows data for a 7\% He$^+$ concentration ($n_1/n_e = 0.072$), which is the same simulation shown in Fig.~\ref{fg:fft} where two-stream instabilities were observed near the sheath. The friction term is the dominant force balancing the electric field near the sheath edge at this concentration. The profile of the maximum linear growth rate calculated from Eq.~(\ref{eq:ephat}) using the midpoint concentration and temperatures along with the profile of $\Delta V$ from the simulated ion speeds is also shown. Here, the friction term is strongly correlated with the instability growth rate, in accordance with the prediction of IEF. Since no explicit collision routine was included for Coulomb collisions, and since ion-neutral collision were turned off, all of the observed ion friction must arise from instability-driven fluctuations of the electric field. Panel (b) shows data for a 90\% He$^+$ concentration ($n_1/n_e = 0.9$), which is a situation where instabilities were not observed. Here, we see that the electric field and kinetic terms directly balance, and the friction force density is negligible.  This is further evidence that the friction force observed in the simulations is directly associated with the presence of the observed instabilities. 

\subsection{Ion Flow Speeds at the Sheath Edge} 

\begin{figure}
\begin{center}
\includegraphics[width=8cm]{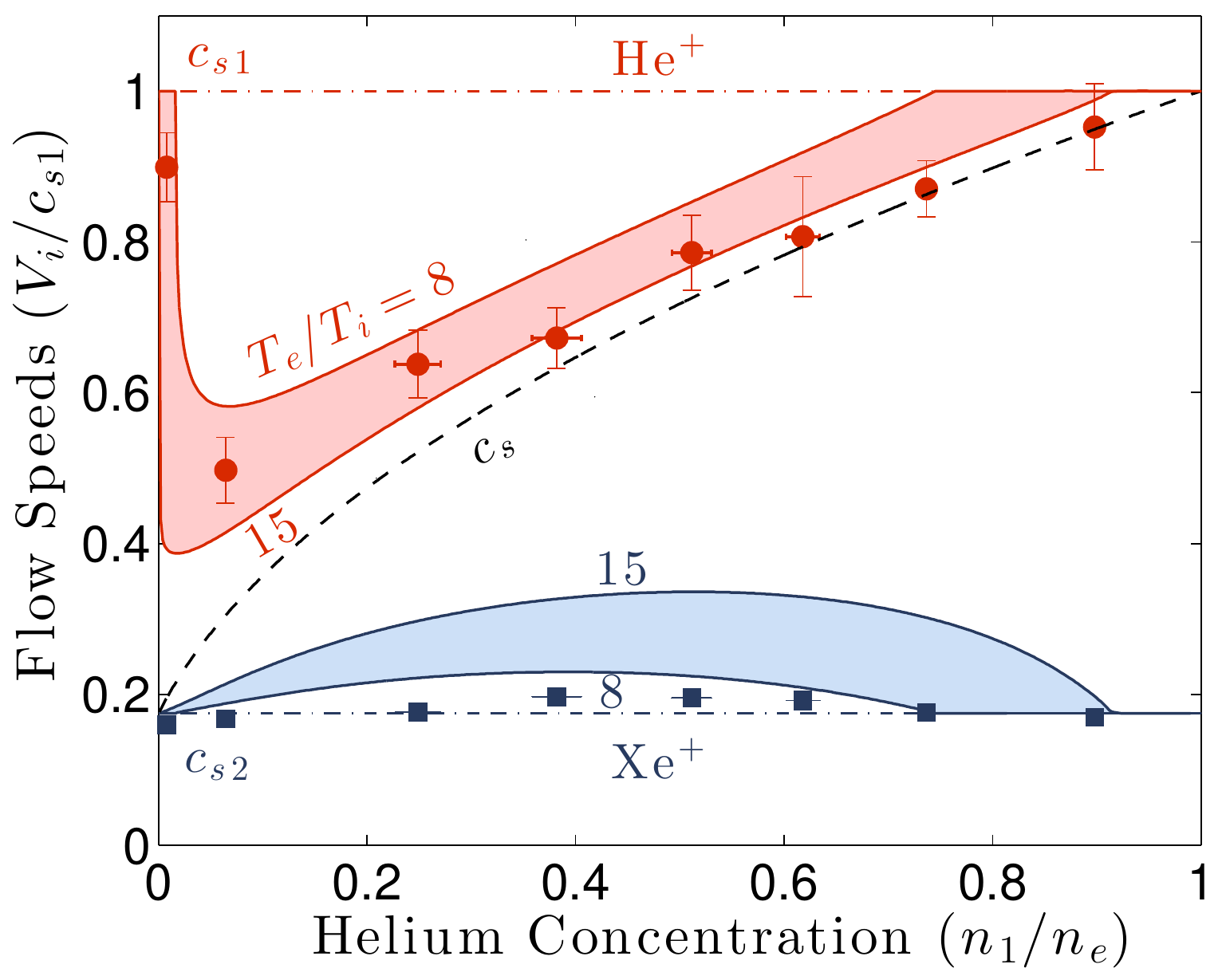}
\caption{He$^+$ (red) and Xe$^+$ (blue) ion flow speeds at the sheath edge. Data points are from PIC-MCC simulations, and solid lines are theoretical predictions of the IEF theory using two limiting values for the electron-ion temperature ratio: $T_e/T_i = 8$ and $15$. The system sound speed $c_s$ is also shown (dashed line) and the speeds are normalized to $c_{s1}$.}
\label{fg:compare}
\end{center}
\end{figure}

Figure~\ref{fg:compare} shows the ion flow speeds observed in multiple simulations at different ion concentrations. This provides a clear demonstration of the merging of ion speeds predicted by the IEF mechanism. Xenon ions are close to their individual sound speed for these parameters, but the helium ions clearly do not agree with either of the simple solutions that have previously been proposed, which are individual sound speeds~\cite{fran:00} or the system sound speed~\cite{lee:07}. The shaded regions show solutions of the IEF theory using numerical solutions of $\Delta V_c$ from Eq.~(\ref{eq:f}) and temperature ratios ($T_e/T_i$) ranging from 8 to 15, which corresponds to the range of temperature ratios observed in the simulations for this data set. The variation in temperature ratio with concentration is caused in part by variations in the electron temperature that can be understood from particle-balance arguments~\cite{lieb:05}, but also in part by variations in the ion temperature. Over this range of values, the IEF theory agrees well with the simulation data, providing the first quantitative test of the theory using PIC-MCC simulations in a situation where the two-stream instabilities arise. The error bars on the simulation points are determined from uncertainty in the sheath edge location (vertical), which we define as the location where $n_i-n_e$ first goes to zero, and uncertainty in the concentration ratio (horizontal). The largest uncertainties in the theoretical analysis include the neglect of ion temperature corrections to the Bohm criterion in Eq.~(\ref{eq:bohm2}), which are of order $T_i/T_e$~\cite{riem:91}, and from the assumption of Maxwellian ion velocity distribution functions. %

\section{Summary} 

In conclusion, we have found the first evidence in PIC-MCC simulations for two-stream instabilities and an associated IEF force in the presheath of plasmas with two ion species. Using numerical calculations for the instability thresholds we reconciled a previously reported discrepancy between simulations and analytic approximations of the IEF theory. This disagreement arose from an assumption of large electron-to-ion temperature ratio in the approximations, but all previous simulations and experiments agree with the instability bounds of the numerical calculation. We also tested these predictions against several new PIC-MCC simulations under a variety of conditions and found good agreement with the theory. 

\section*{Acknowledgments}

This research was supported in part by the University of Iowa and an appointment to the U.S. DOE Fusion Energy Postdoctoral Research Program administered by ORISE (S.D.B.). K.G.\ was supported in part by DOE grant ER55093. T.L.\ would like to thank Dr.\ Jean-Paul Booth for access to computational resources.



\end{document}